\documentclass[preprint]{aastex}

\usepackage{pstricks}
\usepackage{graphicx}
\usepackage{txfonts}
\usepackage{natbib}
\newcommand{\ca}{\ion{Ca}{2} 854.2 nm}
\newcommand{\icore} {$I_{\mathrm{core}}$}
\newcommand{\ibisec} {$I_{\mathrm{bisector}}$}
\newcommand{\acore} {$a_{\mathrm{2}}$}
\newcommand{\acont} {$a_{\mathrm{1}}$}
\newcommand{\ic} {$I_{\mathrm{c}}$}

\begin{document}
\title{Solar cycle variation in Sun-as-a-star \ca\ bisectors }

\author{A. Pietarila and W. Livingston}
\affil{National Solar Observatory, 950 N. Cherry Avenue, Tucson, AZ 85719, USA }

\begin{abstract}
The bisector of the strong chromospheric \ca\ line has an inverse-C
shape the cause of which is not yet fully understood. We show that the
amplitude of the bisector in Sun-as-a-star observations exhibits a
solar cycle variation with smaller amplitudes during highest
activity. The line core intensity is lower during solar minima while
the part of the bisector most sensitive to the line core shows no
systematic change with activity. Our results support the use of
\ca\ bisectors in studying the relationship between convection and
magnetic fields, not only in the Sun but in other stars as well.
\end{abstract}
\keywords{Sun: activity, chromosphere, Line: profiles}

\section{Introduction}

In the Sun photospheric iron lines are on average blue-shifted and
have C-shaped bisectors caused by granular convective motions: there
is an area asymmetry between the up- and downflows as well as a
correlation between intensity and flows, i.e., the upward moving
granules are brighter and larger than the dark downflowing
intergranules (e.g. {\citealt{Beckers+Nelson1978,Nordlund1980}). In
  contrast, the chromospheric \ca\ line exhibits an inverse-C shaped
  bisector and a strong red asymmetry in the line core
  \citep{Uitenbroek2006}. The cause of the inverse-C shape is not
  fully understood. An inverse C-shape is also seen in photospheric
  line bisectors from stars in the hotter half of the H-R diagram
  \citep{Gray+Toner1986}. \cite{Gray2010} attributes it in hot stars
  to a combination of a steep decline in velocities with height and
  the amount of asymmetry between up- and downflows.

Numerical simulations of the Sun, both with and without magnetic
fields, fail to reproduce the observed inverse-C shape of the
\ca\ line bisector \citep{Leenaarts+others2009,Uitenbroek2006}. In
addition, the simulated line core tends to be narrower and deeper than
what is observed \citep{Leenaarts+others2009}. The latter appears to
be at least partly due to insufficient grid resolution, and thus,
reduced velocity amplitudes in the simulations. \cite{Uitenbroek2006}
proposes that the inverse-C bisectors are not solely due to an area
asymmetry, but also a time asymmetry from shock waves: the change from
a sinusoidal wave to one with a sawtooth pattern caused by the
steepening of the wave implies that instead of spending nearly equal
time in up- and downflows a wave spends more time in the downflow
phase leading to the red asymmetry. Based on the computations of
\cite{Uitenbroek2006}, however, shock waves alone cannot account for
the observed inverse-C shape. The 3-dimensional radiative
magnetohydrodynamic simulations including both convection and shock
waves also fail to reproduce the observed bisectors
\citep{Leenaarts+others2009}.

\cite{Livingston1982} found that the bisector amplitude (span) of the
photospheric \ion{Fe}{1} 525.0 nm line was reduced in magnetic regions
relative to non-magnetic regions. He interpreted this as magnetic
fields inhibiting convection and, thus, leading to reduced bisector
amplitudes. A decrease in photospheric bisector amplitudes was found
during the period from 1976 to 1982, i.e., from solar activity minimum
to maximum consistent with increased magnetic flux reducing granular
convection \citep{Livingston1984}. Based on 30-day average mean
magnetic fluxes and photospheric bisector amplitudes,
\cite{Bruning+LaBonte1985} suggested that the decrease in bisector
amplitude is related to old plage regions rather than to active
regions.  \cite{Livingston+others1999} did not find a correlation in
Solar Fourier Transform Spectrometer (FTS) bisector data and activity
indices; instead they found indications of a 2-3 year lag between the
bisector response and solar activity.

In the stellar context the \ca\ line is used as a proxy for magnetic
activity (e.g.,
\citealt{Linsky+others1979,Montes+others2000,Busa+others2007}). NLTE
calculations of the Ca II infrared triplet lines around 850 nm in stellar models
show that the line central depression is less sensitive to
photospheric parameters than the equivalent width, and if rotational
broadening is taken into account, the line central depression is a
purely chromospheric proxy \citep{Andretta+others2005}. Since the Ca
II infrared triplet lines are included in the spectral region that
will be observed by the upcoming GAIA mission, it is important to
understand in more detail how the \ca\ line changes with magnetic
activity.

As further motivation for this study in a broader context, we note
that line bisector analysis can play an important role in the
detection of exoplanets. The signature of stellar spots in radial
velocity (RV) measurements can be similar to that of exoplanets. By
combining RV measurements with bisector measurements it is possible to
evaluate the contribution of stellar activity to the RV variations
(e.g,
\citealt{Queloz+others2001,Povich+others2001,Dall+others2006,Pont+others2010,Figueira+others2010}). \cite{Lagrange+others2010}
used observations of sunspot distributions over a solar cycle to model
the RV curve of a solar-type star. They find that the RV curve is
highly variable depending on activity level and that there is a
correlation between bisector amplitudes and RV even when multiple
spots are present. Understanding the interplay between spot-induced
variation in RV, particularly as reflected in bisector variation, can
help to distinguish between RV variations that are induced by activity
versus those caused by the presence of exoplanets

In this paper we use Sun-as-a-star \ca\ data covering over 25 years in
time to study the solar cycle variation of the \ca\ line bisector and
intensity. We find that the line core intensity and the portion of the
bisector sensitive to the line wings (formed in the upper
photosphere/lower chromosphere) are correlated with activity, while
the bisector portion sensitive to the very line core and, thus purely
chromospheric, is uncorrelated with activity.

\section{Data}

 The \ca\ line has been used in Sun-as-a-star observations at the
 McMath-Pierce telescope (McM/P, \citealt{pierce1964}) since the
 mid-1980s. The observations were made by scanning with the
 spectrograph grating (see \citealt{Brault1971} for a detailed
 description). The spectrograph is in the double-pass mode which compensates for electronic
 offsets and scattered light. A single element silicon diode was used
 to detect the signal. To integrate the sunlight for the Sun-as-a-star
 (full disk) observations the normal concave image-forming mirror of
 the telescope was replaced with a flat resulting in a pinhole image
 of the Sun on the grating. The main change in the observational setup
 was the upgrade of the grating in 1992; the old (25 cm $\times$ 15
 cm) grating was replaced by a larger one (42 cm $\times$ 32 cm). The
 upgrade lead to some changes in the spectra: the line core intensity
 prior was a factor 1.07 higher than after the change,
 perhaps because of better sampling of the limb. Each intensity
 profile consists of 4096 wavelength points and usually between 10 and
 20 scans were averaged for each profile measurement. Two wavelength
 samplings were used: 5.85 m\AA\ and 9.44 m\AA. The larger sampling
 was used before the change in the grating. In the analysis we include
 profiles with both wavelength samplings.

For the present analysis we use all the full disk \ca\ data,
altogether 1035 profiles, from the time period 1985 to 2002 omitting profiles for which remarks of bad seeing or instrumental problems were found in the
observing logs. Thus, we have coverage over solar cycles 22 and
23. There are lengthy periods when no data are available, especially
between 1990 and 1996. However, we have very good coverage between
1996 and 2002 coinciding with the rising phase of cycle 23.

To verify our results we repeat the analysis with the daily synoptic
spectral observations recorded by the SOLIS Integrated Sunlight
Spectrometer (ISS). ISS observes the Sun-as-a-star in various
wavelength regions, including the \ca\ line (design spectral resolution
300000). We use data starting from 2006 (first available ISS data) to
the end of 2010. The range of activity covered by the data set is
limited compared to the McM/P set. We do not directly compare values
of the bisectors from ISS and McM/P, but instead focus on temporal
trends within an instrument.

\section{Results}

We measure the bisectors for each of the 1035 observed McM/P
profiles. Because there is a blend in the blue wing of the \ca\ line,
we only include intensities below $I/$\ic=0.45 (with \ic\ the continuum
intensity) in the bisector analysis. No blends are present in the red
wing. For each measured profile we compute the line core intensity,
\icore, and three parameters to describe the bisector (see
Fig.~\ref{f1}): \ibisec, which is the $I/$\ic\ where the red excursion
of the bisector is the largest, \acore\ is the amplitude of the
bisector's lower part (between \icore and \ibisec) and \acont\ of the
upper portion (between \ibisec\ and $I$/\ic=0.45). The line core is
formed in the chromosphere while the formation of the wings is in the
upper photosphere. Therefore, \acore\ describes the truly
chromospheric component of the bisector and \acont\ includes an upper
photospheric (starting at z=500 km above $\log(\tau_{500\ \mathrm{nm}}) = 0$)
component. By considering only relative measures, such as bisector
amplitudes (distances between two wavelength points), instead of
parameters requiring an absolute wavelength calibration, such as the
line core position, we avoid uncertainties due to the determination of
the zero wavelength.

\begin{figure*}
\includegraphics[width=10cm,angle=90]{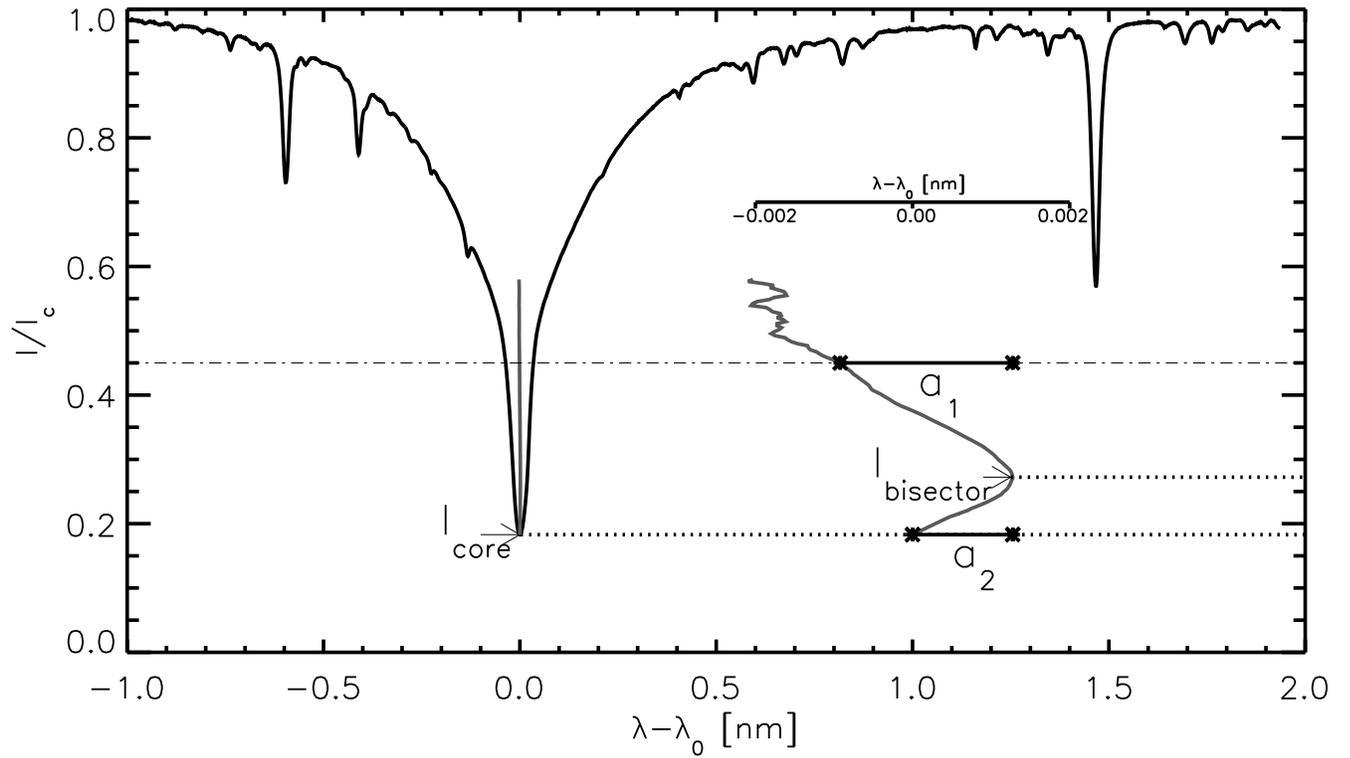}
\caption{Example of a \ca\ profile (black) and its bisector (gray)
  observed on May 7, 1986. The bisector is shown drawn to scale and on
  right expanded by factor 350(corresponding y-axis is shown above the bisector) to illustrate the parameters measured to
  characterize the bisector. The thick horizontal lines mark the
    bisector spans, \acore\ and \acont, the dotted lines mark the
    intensities \ibisec\ and \icore.   }
\label{f1}
\end{figure*}

Fig.~\ref{f2} shows how the \ca\ line changes with the solar
cycle. Three of the measured parameters, \icore, \ibisec, and \acont,
show a solar cycle variation. This is especially well seen during the
rising phase of cycle 23: the core intensity increases as the solar
activity increases, i.e, the line becomes brighter. Also
\ibisec\ increases with activity. \acont\ decreases with increasing
activity while \acore\ shows no systematic temporal trend.

\begin{figure*}
\begin{center}
\includegraphics[width=14cm]{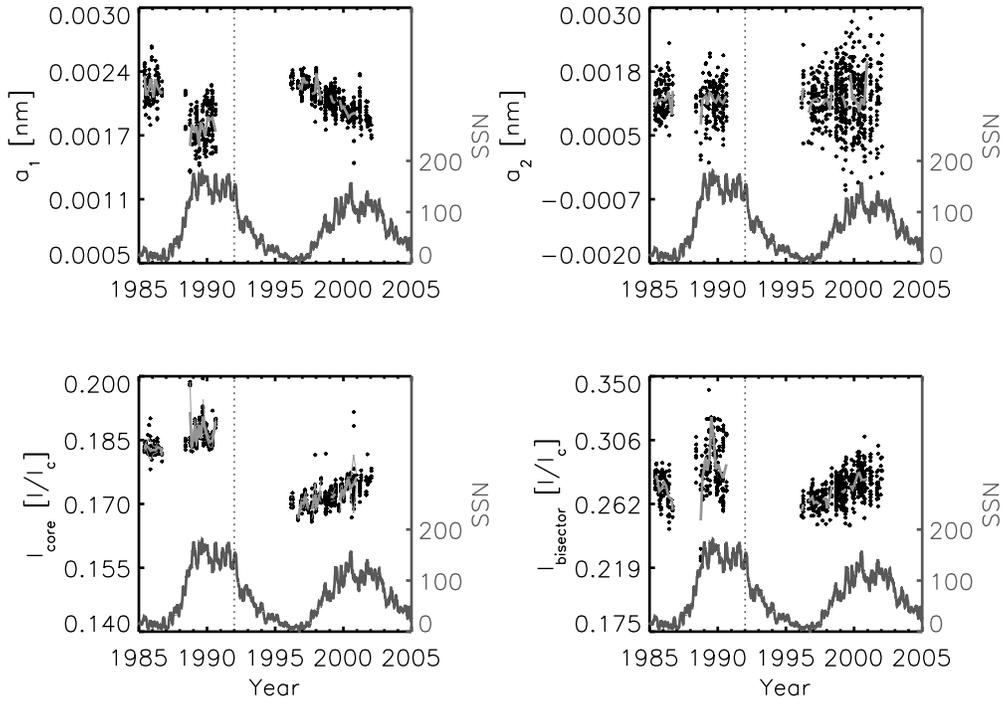}
\caption{\ca\ line through the solar cycle. Top row: \acont,
  \acore. Bottom row: \icore, \ibisec. The over-plotted thin gray line
  is the 2-month mean. In gray: the 54-day running average of the
  daily sunspot number. The vertical dotted line marks the change in
  spectrograph grating.}
\label{f2}
\end{center}
\end{figure*}

\begin{table}
  \caption{Correlation between \ca\ line parameters and SSN.}
  \label{table:corr}      
  \centering                          
  \begin{tabular}{ll}
    \hline\hline
     Parameter &  $r_{c}$ \\
    \hline    
    {\bf McM/P}: & \\
    \acont & -0.67 (-0.72) \\
    \acore & -0.071 \\
    \icore & 0.5 \\
    \ibisec & 0.46\\
    {\bf ISS}: & \\
    $a_{bisector}$ & -0.46\\

    \hline
  \end{tabular}
\end{table}

Of the two measures of the bisector amplitude, only \acont\ (the
component closer to continuum) is clearly correlated with the daily
sunspot number (SSN, Fig.~\ref{f3}). Of all the parameters measured
here, it has the strongest anticorrelation with SSN: the Pearson
correlation coefficient is -0.67. See Table~\ref{table:corr} for
 a summary of the correlation coefficients. If we only include
profiles with the smaller wavelengths sampling, 5.85 m\AA, the
correlation coefficient for \acont\ and SSN is slightly larger:
-0.72. This may be because of the better sampling or because the
observations with the new grating coincide with the fairly monotonic
rise of cycle 23. No correlation is seen for SSN and
\acore\ (r=-0.071). The correlation coefficient (after applying
scaling factor 1.07 to account for the grating change) for SSN and
\icore\ is 0.50 and for SSN and \ibisec\ 0.46.

Variation of \icore\ with solar activity is a possible cause for the
solar cycle dependence of \acont. To rule this out, we recompute \acont: instead of using a fixed $I/$\ic=0.45 to define
the intensity range from which the bisector amplitude is computed, we
redefine the range as: [\icore, \icore+$\Delta$I]. The resulting
correlation coefficients for \acont\ and SSN for $\Delta I$=0.2 and
0.3 are -0.54 and -0.67.

To test if there is a temporal delay between SSN and \acont\ we used
the discrete correlation function \citep{Edelson+Krolik1988}. The
results show no evidence of a time lag, instead the function peaks at
0 days (r=-0.24).

The ISS \ca\ data confirm that the change of the bisector amplitude is
related to solar activity and not an artifact in the McM/P data
(Fig.~\ref{f3}). The ISS spectral sampling is not as high as in the McM/P data, so for the ISS profiles we compute a single bisector
amplitude, i.e., we do not divide
the bisector in upper and lower parts. This amplitude essentially measures the same quantity as \acont\ for the McM/P data. The correlation coefficient is
-0.46, clearly lower than for the McM/P data. The range of solar
activity covered by the ISS data is much smaller than in the McM/P
data: the ISS data cover the very quiet solar cycle minimum when the
sunspot number stayed below 50.

\begin{figure*}
\begin{center}
\includegraphics[width=12cm]{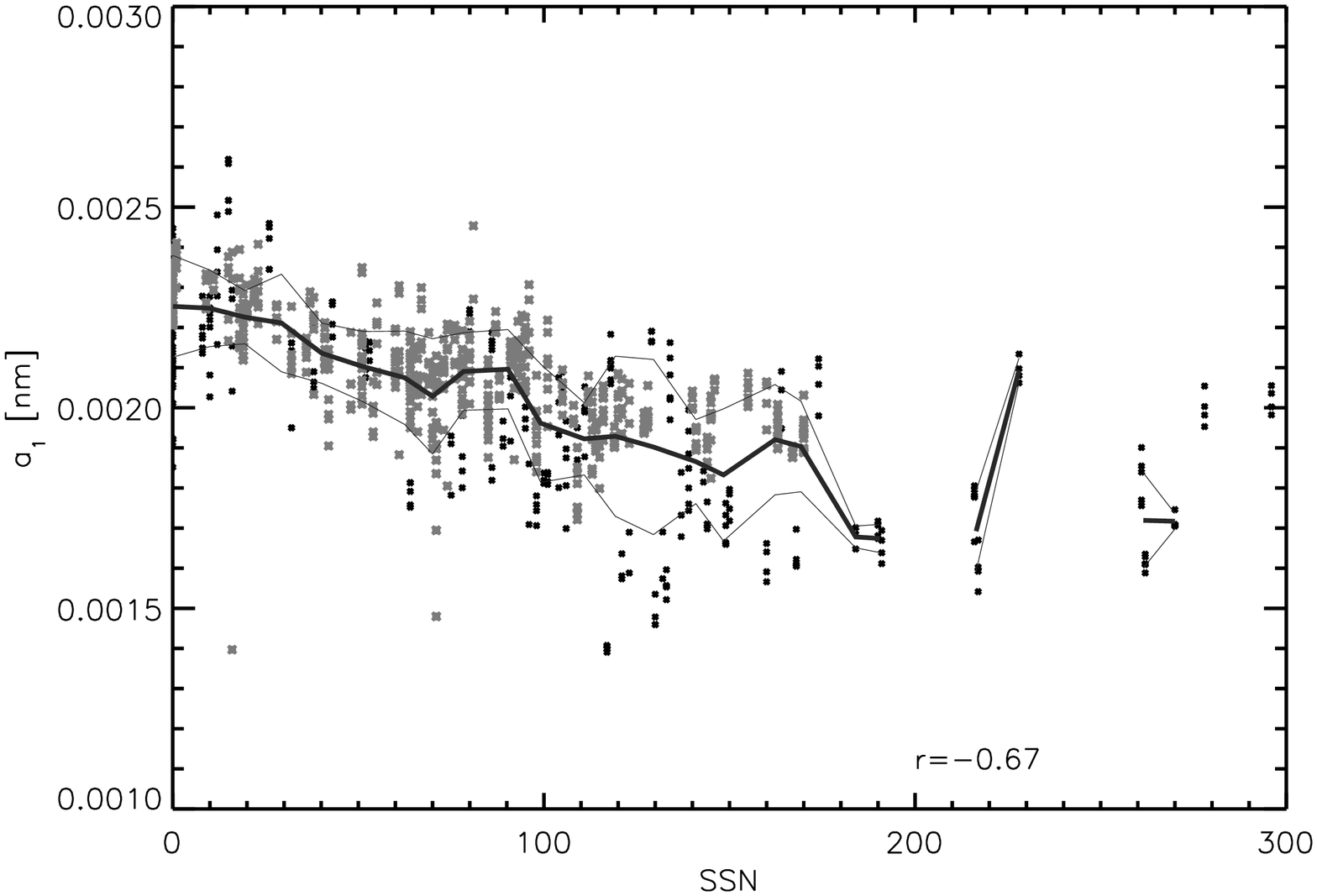}
\includegraphics[width=12cm]{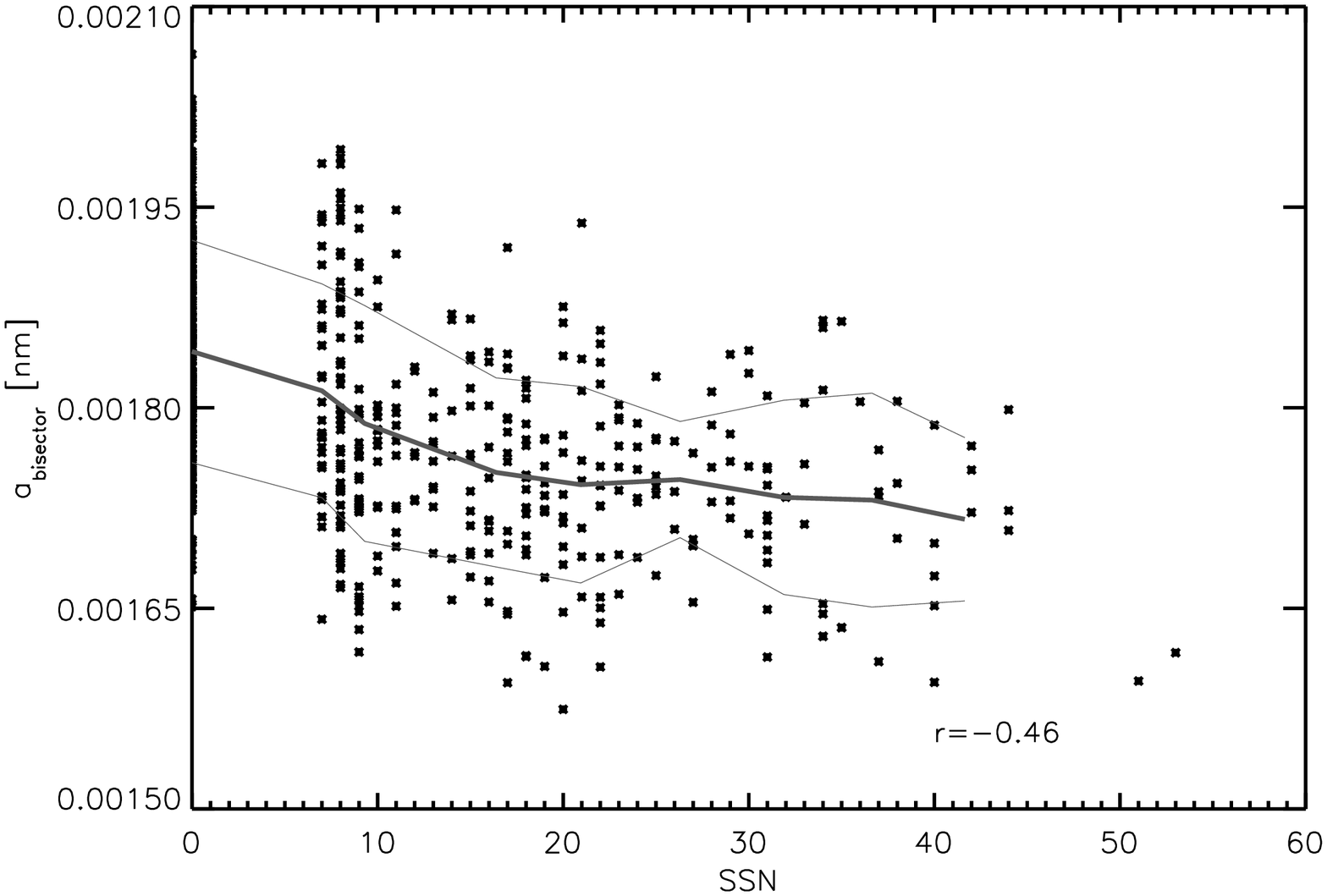}
\caption{McM/P \acont\ (top) and ISS (bottom) \ca\ bisector amplitude
  as a function of the daily sunspot number. The over-plotted lines
  show the mean $\pm$ standard deviation in bins of 10 SSN. The
  Pearson correlation coefficient is given in the lower right
  corner. McM/P data taken after grating change is shown in gray
  asterisk. }
\label{f3}
\end{center}
\end{figure*}

We also computed the correlation between the McM/P bisector amplitudes
and the 1 \AA\ \ion{Ca}{2} K emission index from the NSO/AFRL/Sac peak
Ca K-line Monitoring program \citep{SacPeakCaK}. The correlation
coefficient is -0.40. It should be noted that \ion{Ca}{2} K data are
available for only 420 days of the entire McM/P data set. The ISS also
produces daily 1 \AA\ \ion{Ca}{2} K emission indices. For the ISS
\ca\ bisector amplitude and \ion{Ca}{2} K index data the correlation
is -0.37.

\section{Discussion}

Our analysis combines data from McM/P and ISS and shows that there is a
clear solar cycle variation in the line core intensity and the
bisector of the \ca\ line. Furthermore, the bisector variation is
mostly due to the upper photospheric/lower chromospheric component
(\acont). No solar cycle dependence is seen in the purely
chromospheric bisector component (\acore). The change in
bisector amplitude as a function of sunspot number is seen both in
McM/P and ISS data. The range of sunspot numbers in ISS data is
smaller and also the scatter is larger. Since we see the trend in
both, the formation of the asymmetries in the \ca\ line appears to be
sensitive to fairly small changes in activity. Somewhat weaker
correlations are found for the bisector amplitudes and the 1 \AA\
\ion{Ca}{2} K index.

An inverse correlation between solar activity and \ca\ bisector
amplitudes is not surprising if convection is the cause of the
inverse-C bisectors. Since we only see the variation in \acont\ and
not in \acore, this may indicate that the amplitude of the inverse-C
shape is mainly due to convective overshoot (similar to the regular
C-shaped bisectors) and the lower portion of the bisector,
\acore\ (the purely chromospheric component), is produced by chromospheric dynamics which do not show any clear solar cycle variation when measured this way. Since we made no absolute wavelength calibration of the
data, we cannot say if there is a solar cycle variation in the line
core position. Convection simulations, both with and without magnetic
fields, fail to reproduce the inverse-C shape
\citep{Leenaarts+others2009,Uitenbroek2006}, making convection
unlikely to be the sole explanation.

If instead the inverse-C bisectors are caused by shock waves as
proposed by Uitenbroek (2006), then the amount of energy channeled
into the chromosphere due to shock waves may depend on solar
activity. The reduced bisector amplitude during activity may be
related to the increased area of circumfacular regions, in which the
\ca\ line is on average more symmetric and deeper
\citep{Harvey2005}. The magnetic canopy, where the plasma-$\beta$
($p_{gas}/p_{mag}$) is of the order of unity, is a region where waves
are coupled and may be converted from one type to another
\citep{Bogdan+others2003}. Near active regions the canopy is located
at a lower height and the acoustic waves may be converted before they
are steep enough to produce the observed red asymmetry (Uitenbroek 2006). As pointed out
by Uitenbroek (2006), the inverse-C shape may be produced by a
combination of time and area asymmetries although simulations
including both have so far not been able to reproduce the observed
bisectors. Whether more realistic simulations with higher spatial
resolution will solve the problem remains to be seen.

We find that the bisector amplitude is correlated to the magnetic
activity even during the last extended solar minimum. If canopies are
a factor in producing the observed bisector solar cycle variation,
then the \ca\ line bisector may provide an observational means to
study convection and chromospheric magnetic fields in the Sun as well
as other stars.

The Na I D$_{1}$ line also has a bisector with an inverse-C shape
\citep{Uitenbroek2006}. No long term monitoring of its bisector has
been made, but it would not be surprising if it too exhibited similar
solar cycle variation as \ca. Since the Na I D$_{1}$ line is formed
lower in the atmosphere than \ca, it would be interesting to see if
the line core component of the Na I D$_{1}$ bisector varies with solar
activity. A study of bisectors in lines formed at different heights in
the photosphere and chromosphere might reveal where and how a
transition from the C- to the inverse C-shaped bisectors takes place.

Unlike \cite{Livingston+others1999}, we find no temporal lag in the
bisectors' response to solar activity. It should be noted that
\cite{Livingston+others1999} studied bisector amplitudes of
photospheric lines whereas \ca\ is chromospheric. In \ca\ it is
\acont, not \acore, which varies with solar activity. If the lag in
photospheric line bisector amplitudes is real (the data used by
\cite{Livingston+others1999} have some issues due to imperfect
telescope alignment), a comparison between photospheric and
chromospheric bisector amplitudes could help establish the cause of
the observed inverse-C shaped bisectors.

We plan to expand the study of line bisector amplitudes' solar cycle
dependence to a larger number of solar spectral lines. The ISS data
include a large variety of lines, both chromospheric and photospheric,
which provide a unique data set to study the connection between
convection and magnetic fields. Also of interest is to study in more
detail how the bisectors respond to not only sunspots but also plage,
network fields and canopies. In the stellar context it will be
interesting to characterize how the \ca\ bisector is related to
activity in other stars.

\acknowledgements{The authors are grateful for M. Giampapa and M. Penn
  for helpful comments and discussions regarding this work. The
  National Solar Observatory (NSO) is operated by the Association of
  Universities for Research in Astronomy, Inc., under cooperative
  agreement with the National Science Foundation. SOLIS data used here
  are produced cooperatively by NSF/NSO and NASA/LWS.}

\end{document}